\input{aipcheck}

\documentclass[
    ,final            
  ]
  {aipproc}

\layoutstyle{6x9}

\usepackage{amsmath}

\begin{document}

\newcommand{\etal}{{\it et al.}}
\makeatother

\newcommand{\bra}[1]{\left\langle #1 \right|}
\newcommand{\brared}[1]{\langle #1 ||}
\newcommand{\ee}{\eta^{\ast},\eta}
\newcommand{\product}[2]{\left\langle #1 | #2 \right\rangle}

\newcommand{\kbar}{\bar{k}}
\newcommand{\ket}[1]{\left| #1 \right\rangle}
\newcommand{\ketred}[1]{|| #1 \rangle}
\newcommand{\ahat}{\hat{a}}
\newcommand{\adag}{a^{\dagger}}
\newcommand{\ahatdag}{\hat{a}^{\dagger}}
\newcommand{\Ahat}{\hat{A}}
\newcommand{\Adag}{A^{\dagger}}
\newcommand{\Ahatdag}{\hat{A}^{\dagger}}
\newcommand{\Atdag}{\hat{A}^{(\tau)\dagger}}
\newcommand{\Bdag}{\hat{B}^{\dagger}}
\newcommand{\Bhat}{\hat{B}}
\newcommand{\Bhatdag}{\hat{B}^{\dagger}}
\newcommand{\Btdag}{\hat{B}^{(\tau)\dagger}}
\newcommand{\bhat}{\hat{b}}
\newcommand{\bdag}{b^{\dagger}}
\newcommand{\cdag}{c^{\dagger}}
\newcommand{\chat}{\hat{c}}
\newcommand{\chatdag}{\hat{c}^{\dagger}}
\newcommand{\degree}{^{\circ}}
\newcommand{\sprime}{s^{\prime}}
\newcommand{\Hhat}{\hat{H}}
\newcommand{\Hhatp}{\hat{H}^{\prime}}
\newcommand{\Ihat}{\hat{I}}
\newcommand{\Jhat}{\hat{J}}
\newcommand{\hhat}{\hat{h}}
\newcommand{\fp}{f^{(+)}}
\newcommand{\fpp}{f^{(+)\prime}}
\newcommand{\fm}{f^{(-)}}
\newcommand{\Fhat}{\hat{F}}
\newcommand{\Fhatdag}{\hat{F}^\dagger}
\newcommand{\Fhatp}{\hat{F}^{(+)}}
\newcommand{\Fhatm}{\hat{F}^{(-)}}
\newcommand{\Fhatpm}{\hat{F}^{(\pm)}}
\newcommand{\Fhatdagpm}{\hat{F}^{\dagger(\pm)}}
\newcommand{\Hc}{{\cal H}}
\newcommand{\Hcp}{{\cal H}^{\prime}}
\newcommand{\Ic}{{\cal I}}
\newcommand{\It}{\widetilde{I}}
\newcommand{\ITV}{{\cal I}_{\rm TV}}
\newcommand{\Jc}{{\cal J}}
\newcommand{\jp}{j^{\prime}}
\newcommand{\Qc}{{\cal Q}}
\newcommand{\Pc}{{\cal P}}
\newcommand{\Ec}{{\cal E}}
\newcommand{\Sc}{{\cal S}}
\newcommand{\Rc}{{\cal R}}

\newcommand{\ddg}{d^{\dagger}}

\newcommand{\Nhat}{\hat{N}}
\newcommand{\Nt}{\widetilde{N}}
\newcommand{\Vt}{\widetilde{V}}
\newcommand{\nL}[1]{n_{L_{#1}}}
\newcommand{\nK}[1]{n_{K_{#1}}}
\newcommand{\nKb}{\mbox{\boldmath $n_K$}}
\newcommand{\nLb}{\mbox{\boldmath $n_L$}}

\newcommand{\mubar}{\bar{\mu}}

\newcommand{\Dc}{{\mathscr D}}
\newcommand{\Ddag}{\hat{D}^{\dagger}}
\newcommand{\dhat}{\hat{d}}
\newcommand{\Dhat}{\hat{D}}
\newcommand{\Ghat}{\hat{G}}
\newcommand{\Glambda}{G^{(\lambda)}}
\newcommand{\Gstarlambda}{G^{(\lambda)\ast}}
\newcommand{\Qhat}{\hat{Q}}
\newcommand{\Rhat}{\hat{R}}
\newcommand{\Phat}{\hat{P}}
\newcommand{\Pdag}{\hat{P}^{\dagger}}
\newcommand{\Psihat}{\hat{\Psi}}
\newcommand{\Qdag}{Q^{\dagger}}
\newcommand{\That}{\hat{\Theta}}
\newcommand{\Thatt}{\widetilde{\hat{\Theta}}}
\newcommand{\Tr}{{\rm Tr}}

\newcommand{\ktilde}{\tilde{k}}

\newcommand{\Pcirc}{\stackrel{\circ}{P}}
\newcommand{\Qcirc}{\stackrel{\circ}{Q}}
\newcommand{\Ncirc}{\stackrel{\circ}{N}}
\newcommand{\Tcirc}{\stackrel{\circ}{\Theta}}
\newcommand{\Pcircp}{\stackrel{\circ}{P^{\prime}}}
\newcommand{\Qcircp}{\stackrel{\circ}{Q^{\prime}}}
\newcommand{\Ncircp}{\stackrel{\circ}{N^{\prime}}}
\newcommand{\Tcircp}{\stackrel{\circ}{\Theta^{\prime}}}
\newcommand{\Fp}{F^{(+)}}
\newcommand{\Fm}{F^{(-)}}
\newcommand{\Rp}{R^{(+)}}
\newcommand{\Rm}{R^{(-)}}
\newcommand{\Bt}{\widetilde{B}}
\newcommand{\lambdat}{\widetilde{\lambda}}
\newcommand{\Phatt}{\widetilde{\hat{P}}}
\newcommand{\ab}{\bf a}

\newcommand{\Ab}{\mbox{\boldmath $A$}}
\newcommand{\Abdag}{\mbox{\boldmath $A$}^{\dagger}}
\newcommand{\Bb}{\mbox{\boldmath $B$}}
\newcommand{\cb}{\bf c}
\newcommand{\Db}{\mbox{\boldmath $D$}}
\newcommand{\Nb}{\mbox{\boldmath $N$}}
\newcommand{\Nbhat}{\hat{\mbox{\boldmath $N$}}}
\newcommand{\Qb}{\mbox{\boldmath $Q$}}
\newcommand{\Qhatt}{\widetilde{\hat{Q}}}
\newcommand{\Pb}{\mbox{\boldmath $P$}}
\newcommand{\phit}{\phi(t)}
\newcommand{\pdot}{\dot{p}}
\newcommand{\phix}[1]{\phi(#1)}
\newcommand{\qdot}{\dot{q}}
\newcommand{\phivib}{\phi(\eta^{\ast},\eta)}
\newcommand{\Ts}{{\cal T}}
\newcommand{\del}{\partial}
\newcommand{\eps}{\epsilon}
\newcommand{\beq}{\begin{equation}}
\newcommand{\beqa}{\begin{eqnarray}}
\newcommand{\eeq}{\end{equation}}
\newcommand{\eeqa}{\end{eqnarray}}
\newcommand{\Yb}{${}^{168}$Yb\ }
\newcommand{\Zhat}{\hat{Z}}
\newcommand{\rhodot}{\dot{\rho}}
\newcommand{\Khat}{\hat{K}}
\newcommand{\Kp}{K^{+}}
\newcommand{\Km}{K^{-}}
\newcommand{\Kz}{K^0}

\newcommand{\lb}{\bf l}
\newcommand{\sbold}{\bf s}

\newcommand{\Lp}{L^{+}}
\newcommand{\Lm}{L^{-}}
\newcommand{\Lz}{L^0}

\newcommand{\Mc}{{\cal M}}
\newcommand{\Mchat}{\hat{\cal M}}

\newcommand{\ddeta}{\frac{\partial}{\partial \eta}}
\newcommand{\ddetastar}{\frac{\partial}{\partial \eta^\ast}}
\newcommand{\etastar}{\eta^\ast}
\newcommand{\ketvib}{\ket{\phi (\etastar, \eta)}}
\newcommand{\bravib}{\bra{\phi (\etastar, \eta)}}
\newcommand{\zhateta}{\hat{z}(\eta)}
\newcommand{\zhat}{\hat{z}}
\newcommand{\oo}{\stackrel{\circ}{O}(\etastar,\eta)}
\newcommand{\oodag}{\stackrel{\circ}{O^{\dagger}}(\etastar,\eta)}
\newcommand{\oodagp}{\stackrel{\circ}{O^{\dagger\prime}}(\etastar,\eta)}
\newcommand{\oop}{\stackrel{\circ}{O^{\prime}}(\etastar,\eta)}
\newcommand{\Odag}{\hat{O}^{\dagger}}
\newcommand{\Ohat}{\hat{O}}
\newcommand{\Uinv}{U^{-1}(\etastar, \eta)}
\newcommand{\Uinvp}{U^{-1}(\etastar,\eta,\varphi,n)}
\newcommand{\U}{U(\etastar, \eta)}
\newcommand{\Up}{U(\etastar,\eta,\varphi,n)}
\newcommand{\etader}{\frac{\del}{\del \eta}}
\newcommand{\etastarder}{\frac{\del}{\del \etastar}}

\newcommand{\fb}{\mbox {\bfseries\itshape f}}
\newcommand{\SB}{\mbox {\bfseries\itshape S}}

\newcommand{\bg}{\beta,\gamma}

\newcommand{\vbar}{\bar{v}}

\newcommand{\Udag}{U^{\dagger}}
\newcommand{\Vdag}{V^{\dagger}}

\newcommand{\Wc}{{\cal W}}
\newcommand{\Wcdag}{{\cal W}^{\dagger}}

\newcommand{\Xhat}{\hat{X}}
\newcommand{\Xdag}{\hat{X}^{\dagger}}

\renewcommand{\thanks}{\footnote}
\newcommand\tocite[1]{$^{\hbox{--}}$\cite{#1}}

\title{Microscopic description of large-amplitude shape-mixing dynamics with local QRPA inertial functions}


\author{Nobuo Hinohara}{
address={Theoretical Nuclear Physics Laboratory, RIKEN Nishina Center,
Wako 351-0198, Japan}
}

\author{Koichi Sato}{
  address={Department of Physics, Graduate School of Science, Kyoto
  University, Kyoto 606-8502, Japan},
  altaddress={Theoretical Nuclear Physics Laboratory, RIKEN Nishina
  Center, Wako 351-0198, Japan}
}

\author{Kenichi Yoshida}{
address={Theoretical Nuclear Physics Laboratory, RIKEN Nishina Center,
Wako 351-0198, Japan}
}

\author{Takashi Nakatsukasa}{
  address={Theoretical Nuclear Physics Laboratory, RIKEN Nishina Center,
Wako 351-0198, Japan}
}

\author{Masayuki Matsuo}{
  address={Department of Physics, Faculty of Science, Niigata
  University, Niigata 950-2181, Japan}
}

\begin{abstract}
We introduce a microscopic approach to derive all the inertial functions in the 
five-dimensional quadrupole collective Hamiltonian.
Local normal modes are evaluated on the constrained mean field
in the quasiparticle random-phase approximation in order to derive the inertial functions.
The collective Hamiltonians for neutron-rich Mg isotopes are 
determined with use of this approach, and the shape coexistence/mixing 
around the $N=20$ region is analyzed.
\end{abstract}

\maketitle


\section{INTRODUCTION}

The five-dimensional (5D) quadrupole collective Hamiltonian is a powerful theoretical approach to
describe and understand the properties of the low-lying 
collective states (for a recent review, see Ref.~\cite{0954-3899-36-12-123101}).
The collective Hamiltonian consists of,
vibrational and rotational kinetic terms, and the model contains six collective 
inertial functions (three vibrational masses and three rotational moments of inertia), in addition to the collective potential.
All the quantities are functions of quadrupole deformation variables $\beta$ and $\gamma$.
The collective inertial functions are 
usually calculated by means of the Inglis-Belyaev (IB)
cranking approximation \cite{Beliaev1961322}, 
which is derived in the adiabatic perturbation treatment of moving mean field.
The most serious shortcoming of the IB approximation is that the time-odd terms of the 
moving mean field are ignored.
To overcome this shortcoming,
adiabatic time-dependent Hartree-Fock-Bogoliubov (ATDHFB) theories
were developed \cite{Baranger1978123,Dobaczewski1981123},
but the inertial functions derived from ATDHFB theories
have not been used in realistic applications so far.
Recently we have formulated the constrained Hartree-Fock-Bogoliubov
plus local quasiparticle random-phase approximation (CHFB+LQRPA)
method \cite{PhysRevC.82.064313} 
to calculate the collective inertial functions on the basis of the 
adiabatic self-consistent collective coordinate (ASCC) method 
\cite{PTP.103.959}.  
This new method enables us to evaluate 
the inertial functions including the time-odd contribution of the moving field. 
It has been successfully applied to 
the oblate-prolate shape coexistence/mixing phenomena in 
proton-rich Se and Kr isotopes \cite{PhysRevC.82.064313, Sato201153}.

In this presentation, the large-amplitude deformation dynamics 
in low-lying states of neutron-rich Mg isotopes is 
discussed with the use of the CHFB+LQRPA method.
The microscopic mechanism of breaking the $N=20$ shell gap 
and nature of deformation in this ``island of inversion'' region is 
currently under active discussion both experimentally and theoretically.
The significant increases of the $E(4^+)/E(2^+)$ ratio 
and the $B(E2;2^+\rightarrow 0^+)$ value from $^{30}$Mg to $^{34}$Mg 
clearly indicate a rapid growth of quadrupole deformation. 
The experimental data suggest a kind of quantum phase transition 
taking place around $^{32}$Mg and stimulate a microscopic investigation on  
large-amplitude collective dynamics unique to this region of nuclear chart. 

\section{CHFB+LQRPA method}

The 5D quadrupole collective Hamiltonian is written as
\begin{align}
 {\cal H}_{\rm coll} = & T_{\rm vib} + T_{\rm rot} + V(\bg), \\
 T_{\rm vib} =& \frac{1}{2}D_{\beta\beta}(\bg)\dot{\beta}^2 + D_{\beta\gamma}(\bg)\dot{\beta}\dot{\gamma}
 + \frac{1}{2}D_{\gamma\gamma}(\bg)\dot{\gamma}^2,\\
 T_{\rm rot} =& \frac{1}{2} \sum_{k=1}^3 {\cal J}_k(\bg) \omega^2_k,
\end{align}
where $V(\bg)$ is the collective potential, $D_{\beta\beta}, D_{\beta\gamma}$, and $D_{\gamma\gamma}$ are the vibrational inertial functions,
${\cal J}_k$ are the rotational moments of inertia about three principal axes.

To derive these functions in the 5D quadrupole collective Hamiltonian
from the microscopic Hamiltonian $\Hhat$, 
we start from the microscopic theory of large-amplitude collective motion,
called the ASCC method \cite{PTP.103.959}.
The ASCC method is a theory to extract a collective subspace which is embedded in the large-dimensional time-dependent HFB (TDHFB) manifold.
For the present purpose of deriving the 5D collective Hamiltonian,
we start from the two-dimensional version of the ASCC method,
in which we suppose the existence of a set of two-dimensional canonical collective variables (coordinates and momenta) $(q^1, q^2)$ and $(p_1, p_2)$ 
describing the vibrational degrees of freedom.
The set of two collective coordinates $(q^1, q^2)$ has a one-to-one correspondence to the quadrupole deformation variable set $(\bg)$.
The CHFB + LQRPA equations are derived from the two-dimensional ASCC 
method making 
practical approximations 
(see Ref.~\cite{PhysRevC.82.064313} for details of the derivation).
The central concept of CHFB + LQRPA method is the local normal modes
built on constrained mean field defined at each point of the $(\bg)$ plane.
The two-dimensional collective space are spanned 
by CHFB states $\ket{\phi(\bg)}$, which are determined by solving 
the CHFB equations
with four constraints on neutron and proton numbers 
and quadrupole deformations.
The LQRPA equations for vibrational degrees of freedom 
are written as 
\begin{align}
 \delta & \bra{\phi(\bg)}[\Hhat_{\rm CHFB}(\bg), \Qhat^i(\bg)]
  - \frac{1}{i} \Phat_i(\bg) \ket{\phi(\bg)} = 0, \\
 \delta & \bra{\phi(\bg)}[\Hhat_{\rm CHFB}(\bg), \frac{1}{i} \Phat_i(\bg)]
  - \omega^2_i(\bg) \Qhat^i(\bg) \ket{\phi(\bg)} = 0.
 \end{align}
Here $\Hhat_{\rm CHFB}$ is the microscopic CHFB Hamiltonian
including four linear constraint terms, and
$\Qhat^i$ and $\Phat_i$ are local infinitesimal generators defined at $(\bg)$
with respect to the CHFB states.
There exist a lot of solutions of the LQRPA equations, 
and we select two LQRPA solutions with the minimal metric criterion;
at each point of the $(\bg)$ plane, we evaluate the vibrational part of the
metric $W(\bg) = D_{\beta\beta} D_{\gamma\gamma} - D_{\beta\gamma}^2$
for all combinations of two LQRPA modes, and find the pair of modes that gives
the minimal value.
The rotational moments of inertia $\Jc_k$ are evaluated 
at each ($\beta,\gamma$) point
using the LQRPA equations for the rotational degrees of freedom:
\begin{align}
\delta \bra{\phi(\bg)}[\Hhat_{\rm CHFB}(\bg), \Psihat_k(\bg)]
 - \frac{1}{i} \{\Jc_k(\bg) \}^{-1} \Ihat_k \ket{\phi(\bg)} = 0,
\end{align}
\begin{align}
 \bra{\phi(\bg)} [\Psihat_k(\bg), \Ihat_{k'} ]\ket{\phi(\bg)} = i \delta_{kk'}.
\end{align}

\section{APPLICATION TO THE NEUTRON-RICH MG ISOTOPES}

\subsection{Details of numerical calculation}

Numerical calculations are performed 
using the pairing-plus-quadrupole (P+Q) Hamiltonian
including the quadrupole-pairing interaction.
The two-major $sd$ and $pf$ shells are taken into account 
for both neutrons and protons as an active single-particle model space.
In order to determine the parameters in the P+Q Hamiltonian,
we have performed HFB calculations 
using the Skyrme SkM* functional 
and  the HFBTHO code \cite{Stoitsov200543}.
In the Skyrme-HFB calculation, 
the volume-pairing strength is adjusted to reproduce the experimental neutron pairing gap of $^{30}$Ne (1.26MeV).
The single-particle energies of the P+Q Hamiltonian
are taken from those obtained in the Skyrme-HFB calculation
after the effective mass scaling.
The strengths of the monopole-pairing interaction 
are adjusted to reproduce 
the Skyrme-HFB result at the spherical configuration.
The strength of the quadrupole particle-hole interaction are adjusted to 
reproduce the magnitude of 
the axial quadrupole deformation at the HFB minimum.
The strength of the quadrupole-pairing interaction is determined with 
the self-consistent prescription \cite{Sakamoto1990321}.
We use the quadrupole polarization charge $\delta e_{\rm pol} = 0.5$ 
for both neutrons and protons  
when evaluating $B(E2)$ and quadrupole moments.

\subsection{Results}

\begin{figure}
\begin{tabular}{cc}
	\includegraphics[width=60mm]{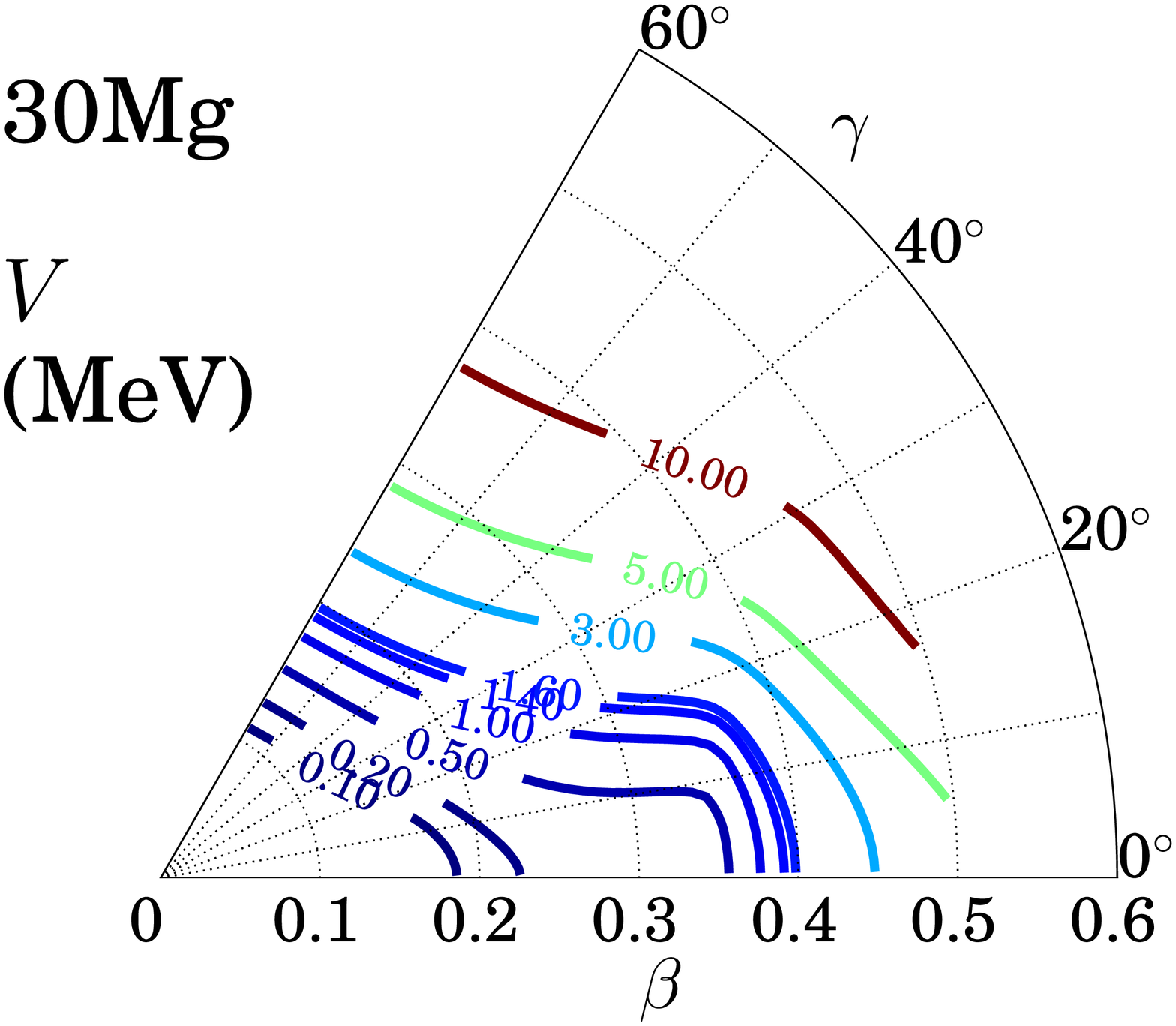} &
	\includegraphics[width=60mm]{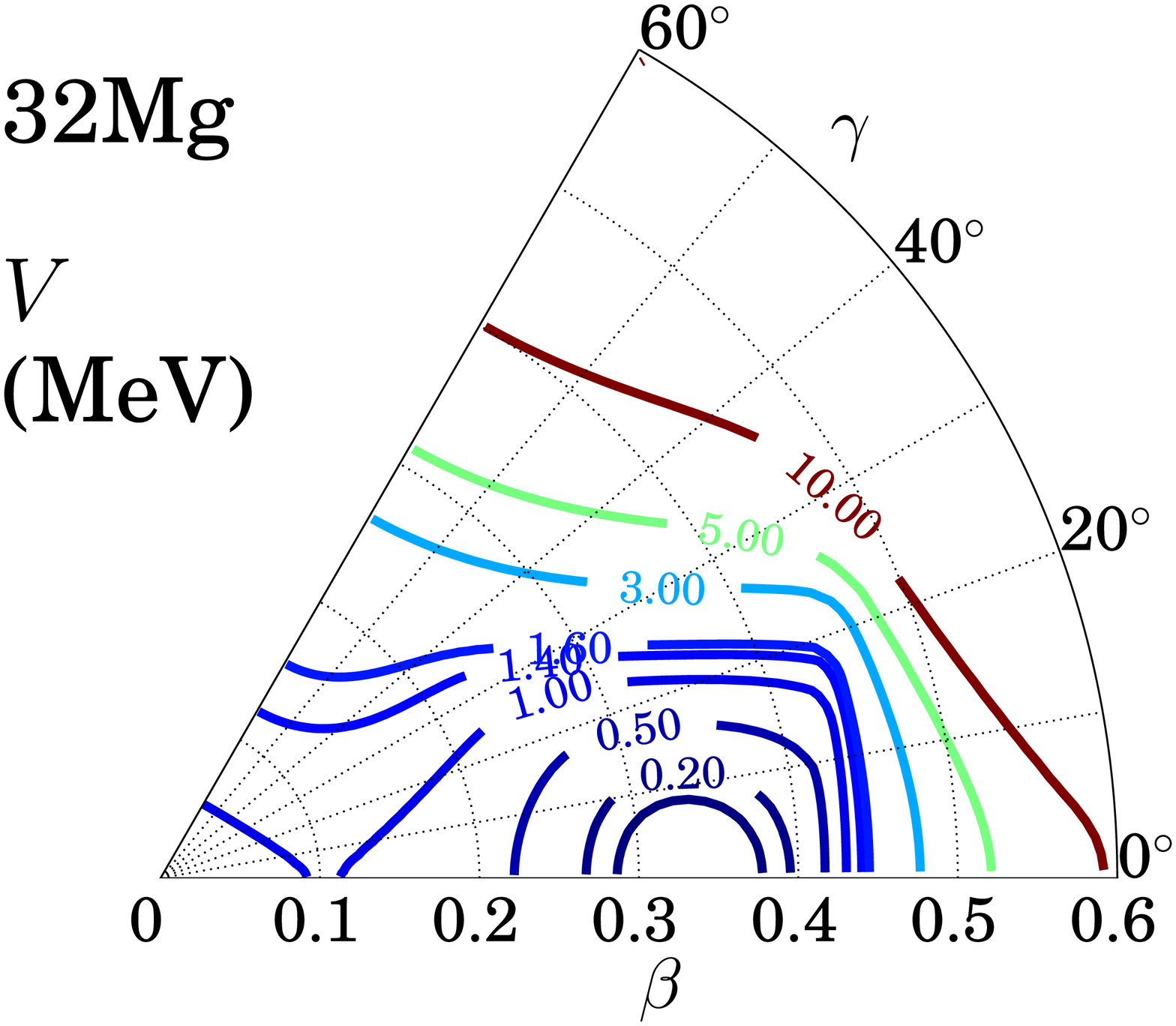} \\
	\includegraphics[width=60mm]{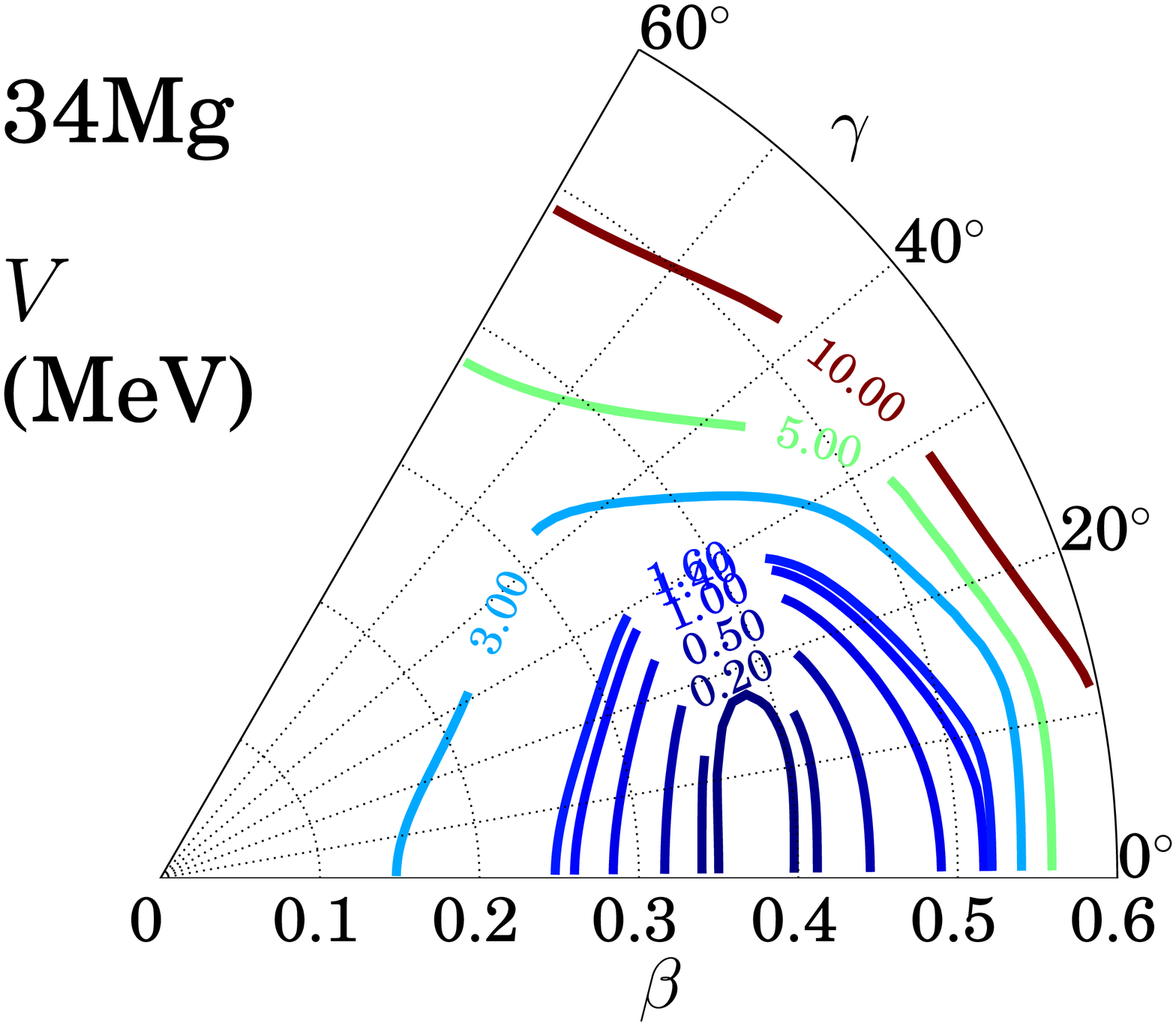} &
	\includegraphics[width=60mm]{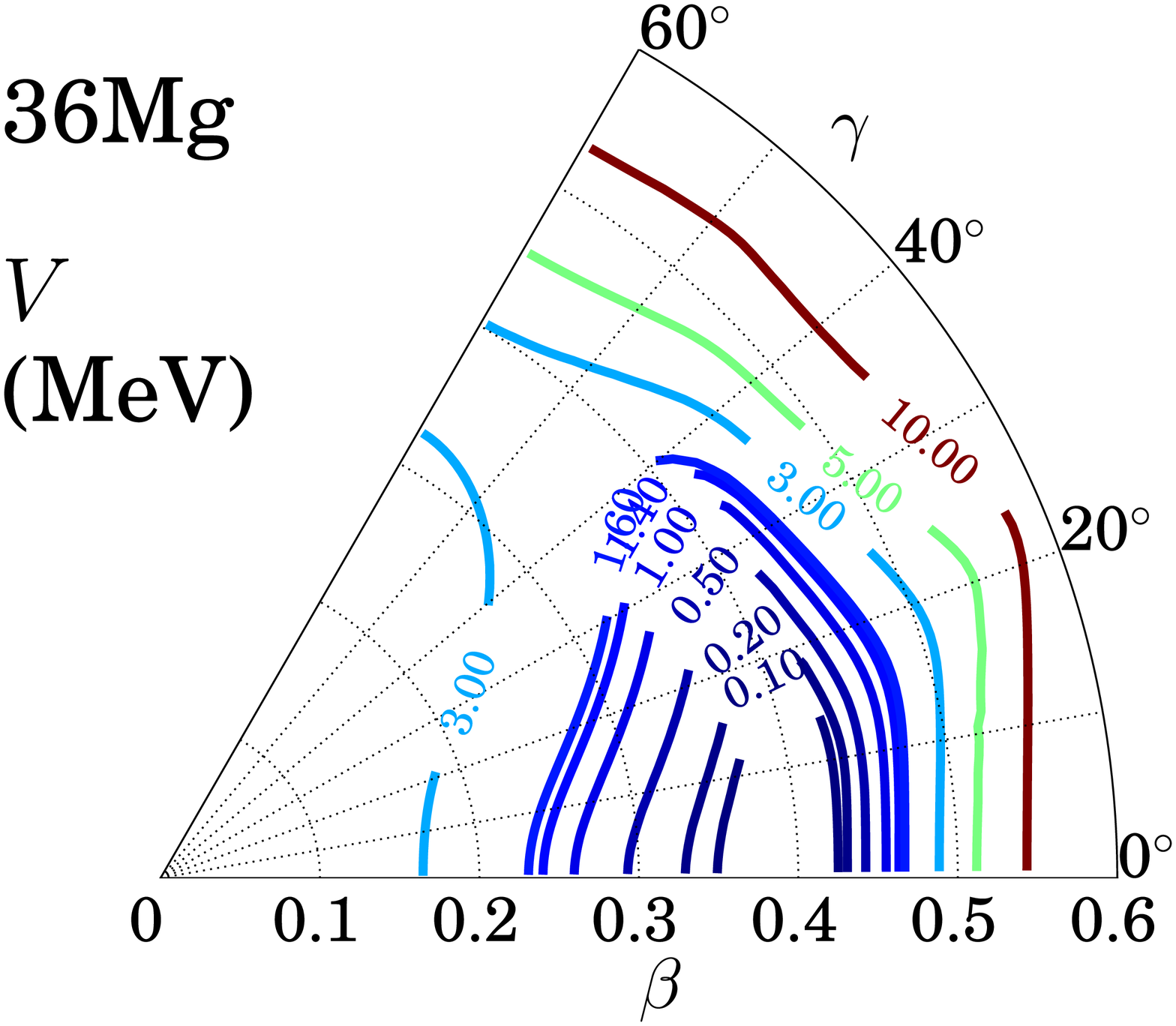}
\end{tabular}
\caption{ \label{fig:V}
Potential energy surfaces of $^{30-36}$Mg.}
\end{figure}

Figure~\ref{fig:V} shows the potential energy surfaces 
for $^{30-36}$Mg determined 
by solving the CHFB equations.
One can clearly see that the prolate deformation 
grows with increasing neutron number.
The potential energy surface for $^{30}$Mg 
is very soft against the axial quadrupole deformation. 
It has a HFB local minimum at $\beta = 0.11$.
The potential energy surface for $^{32}$Mg 
exhibits a spherical and prolate shape coexistence, 
where the spherical shape is associated with the neutron shell gap at $N=20$.
The spherical local minimum disappears in $^{34,36}$Mg, and the prolate minima
become soft in the triaxial direction.

\begin{figure}
\begin{tabular}{cc}
	\includegraphics[width=60mm]{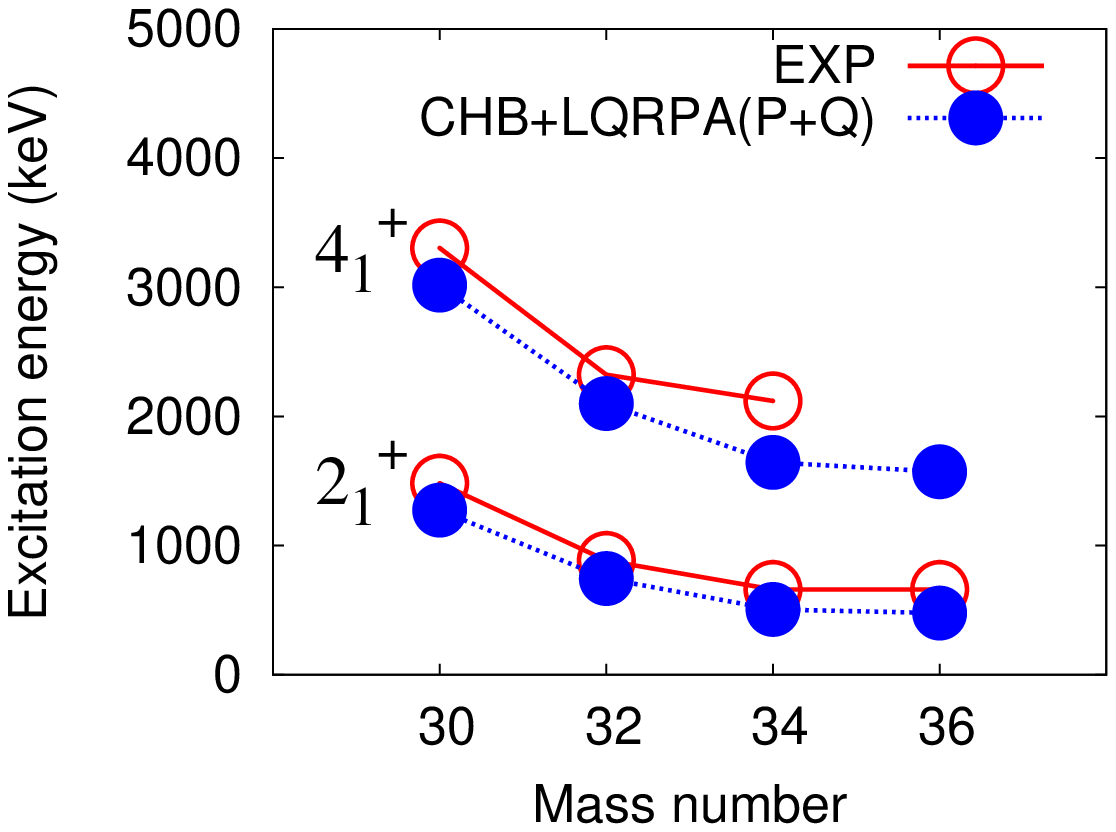} & 
	\includegraphics[width=60mm]{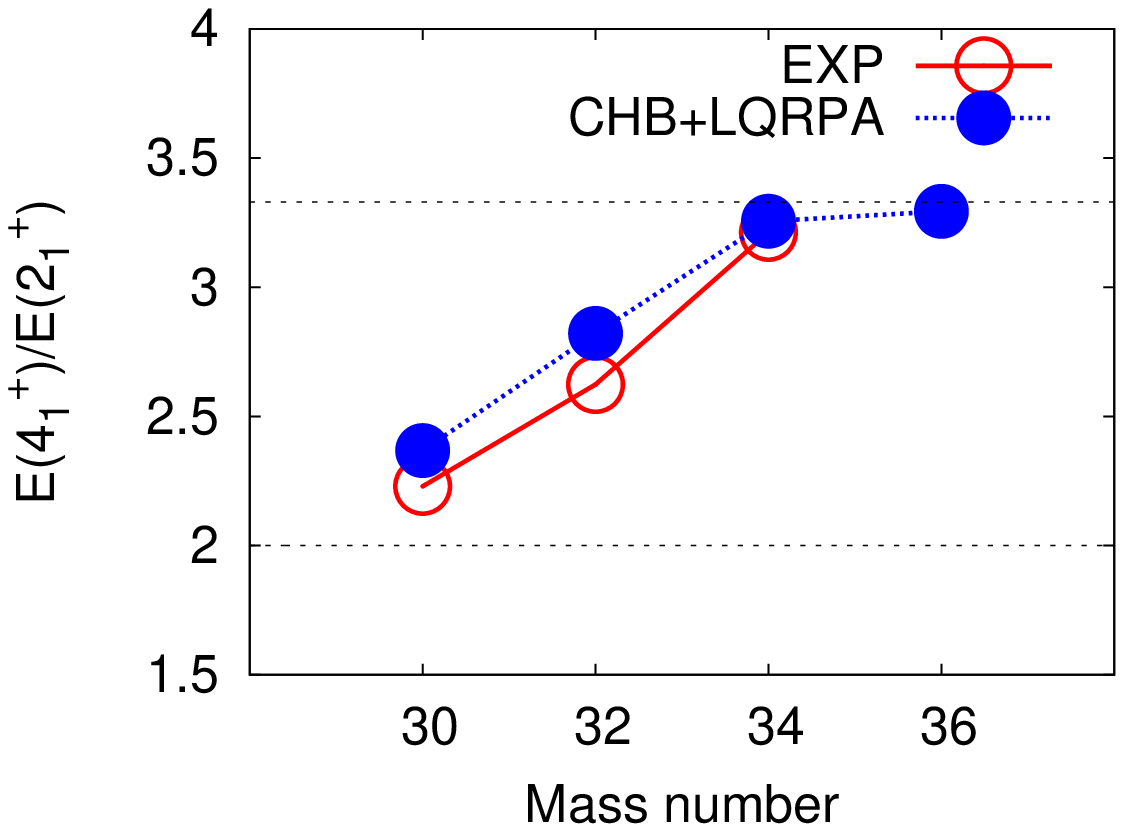} \\
	\includegraphics[width=60mm]{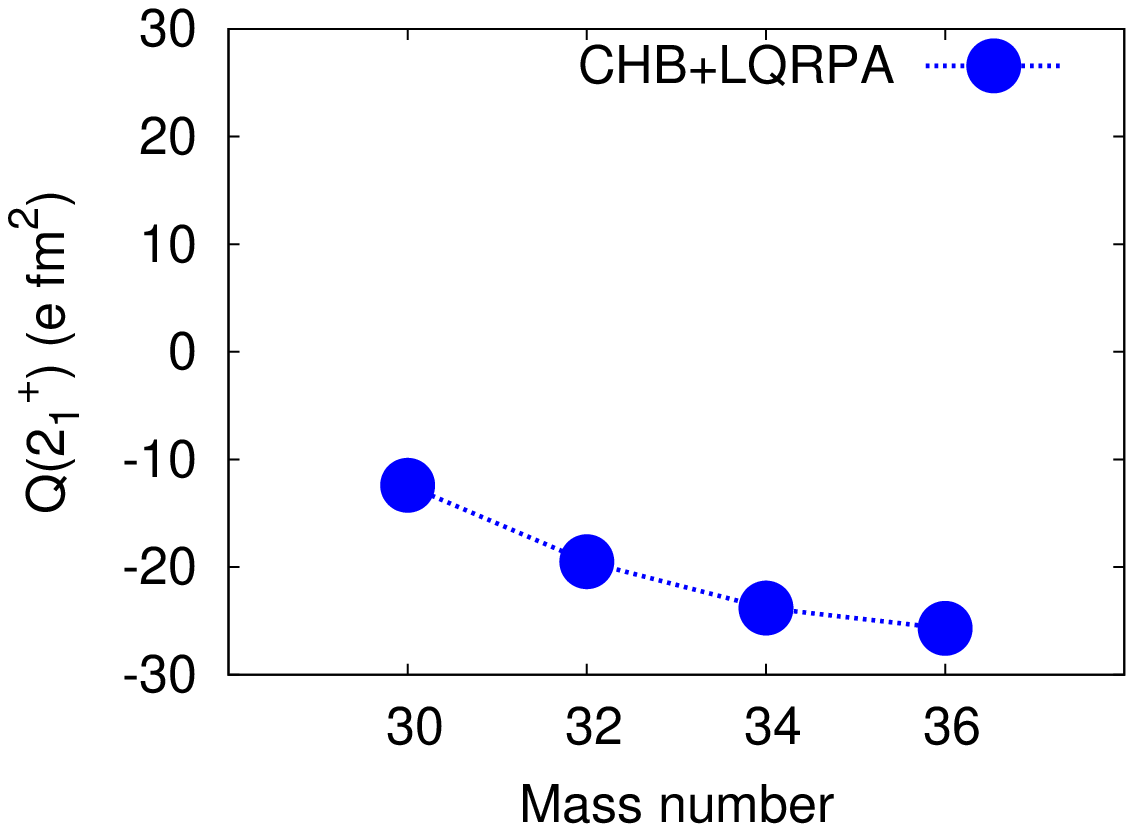} &
	\includegraphics[width=60mm]{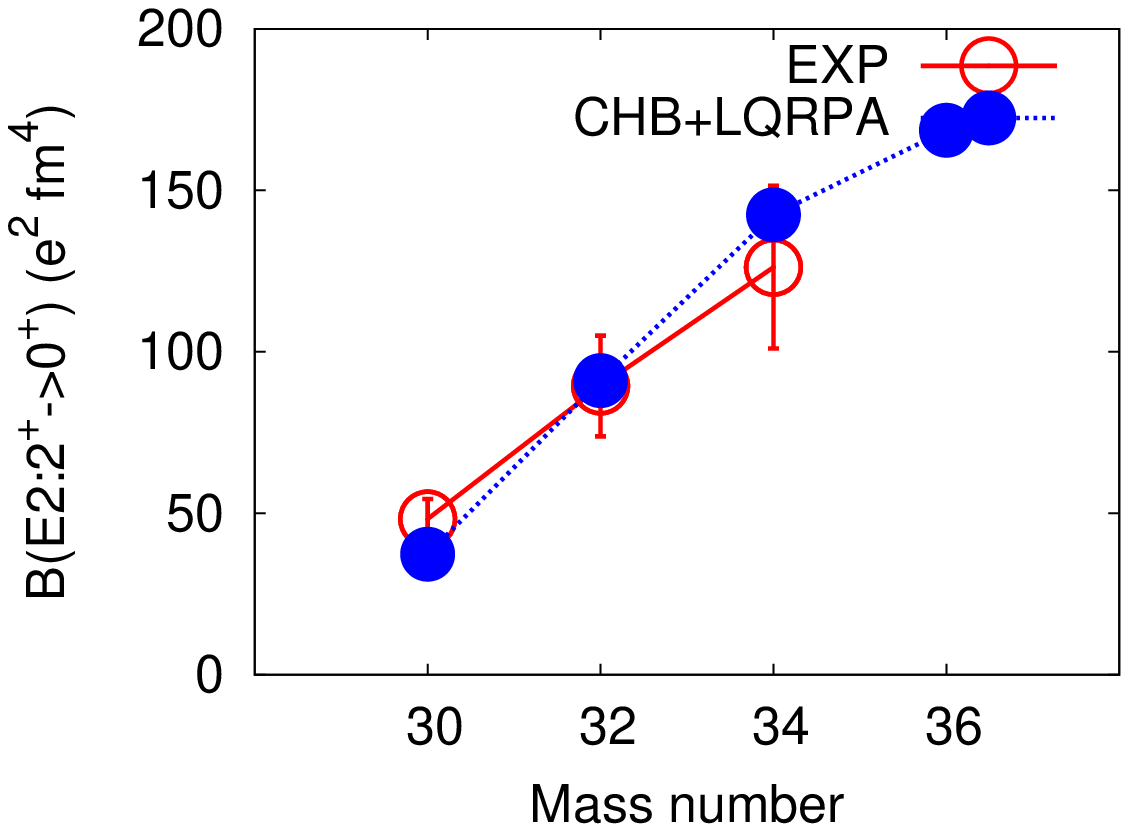}
\end{tabular}
\caption{ \label{fig:energy}
Excitation energies of the $2_1^+$ and $4_1^+$ states, their ratios, 
spectroscopic quadrupole moments $Q$ of the $2_1^+$ state,  
and $B(E2:2_1^+ \rightarrow 0_1^+)$ values calculated for yrast states 
in $^{30- 36}$ Mg.
They are compared with the experimental data \cite{PhysRevC.82.034305,PhysRevC.79.054319,Yoneda2001233,gade:072502,PhysRevLett.94.172501,Motobayashi19959,Iwasaki2001227}.}
\end{figure}

Figure \ref{fig:energy} shows the excitation energies, the spectroscopic 
quadrupole moments, and the $E2$ transition strengths, 
which are obtained by solving the collective Schr\"odinger equation.
The calculation well reproduces the experimental trend,
such as the lowering of the $2_1^+$ energy and 
the increase of the $E(4_1^+)/E(2_1^+)$ ratio from $^{30}$Mg to $^{34}$Mg. 
The behavior of the yrast properties of $^{30-36}$Mg indicates 
that  $^{30}$Mg and  $^{32}$Mg are situated in the center of 
the transitional region of quantum phase transition from spherical to deformed.
In particular, it is very interesting to see that
the potential energy surface of $^{32}$Mg shows 
a spherical-prolate shape coexistence.  
This is a feature somewhat different  
from the shape phase transition known in the heavy-mass region.
The remarkable increase of the $B(E2)$ strength
from $^{30}$Mg to $^{36}$Mg 
agrees with the available experimental data.
More detailed analysis including the properties of the excited bands 
is now in progress.

\section{CONCLUSION}

We have developed a practical microscopic method 
for determining the six inertial functions 
in the five-dimensional quadrupole collective Hamiltonian.  
This method is based on the ASCC method and 
called the CHFB+LQRPA method.  
This new method has been applied to 
the large-amplitude shape-mixing dynamics in the yrast bands of 
neutron-rich Mg isotopes.
The result of numerical calculation for excitation energies 
and $B(E2)$ strengths is in good agreement with the experimental data.
It demonstrates that the present method is powerful to describe the
large-amplitude quadrupole collective motion in the $^{32}$Mg region 
as well as the proton-rich Se and Kr isotopes \cite{PhysRevC.82.064313,Sato201153}.
It also indicates a wide applicability of the CHFB+LQRPA method 
to various kinds of collective dynamics.
More detailed analysis of shape mixing properties in the Mg 
isotopes will be reported in a forthcoming paper \cite{Mginprep}
including the second $0^+$ and $2^+$ states in excited bands.  
Application of this new method with use of modern energy density 
functional is an important future subject.  


\begin{theacknowledgments}
Three of the authors (N.H., K.S., and K.Y.) are supported 
by the Special Postdoctoral Researcher Program (N.H. and K.Y.)
and the Junior Research Associate Program (K.S.) 
of RIKEN.
The numerical calculations were carried out on RIKEN Cluster
of Clusters (RICC) facility.
\end{theacknowledgments}



\bibliographystyle{aipproc}   

\bibliography{../../../../../bibtex/paper}

\begin{thebibliography}{10}
\expandafter\ifx\csname natexlab\endcsname\relax\def\natexlab#1{#1}\fi
\providecommand{\enquote}[1]{``#1''}
\expandafter\ifx\csname url\endcsname\relax
  \def\url#1{\texttt{#1}}\fi
\expandafter\ifx\csname urlprefix\endcsname\relax\def\urlprefix{URL }\fi
\providecommand{\eprint}[2][]{\url{#2}}

\bibitem[Pr\'{o}chniak and Rohozi\'{n}ski(2009)]{0954-3899-36-12-123101}
L.~Pr\'{o}chniak and S.~G. Rohozi\'{n}ski, \emph{J. Phys. G} \textbf{36},
  123101 (2009).

\bibitem[Beliaev(1961)]{Beliaev1961322}
S.~T. Beliaev, \emph{Nucl. Phys.} \textbf{24}, 322(1961).

\bibitem[Baranger and V\'en\'eroni(1978)]{Baranger1978123}
M.~Baranger and M.~V\'en\'eroni, \emph{Ann. Phys.} \textbf{114}, 123
  (1978).

\bibitem[Dobaczewski and Skalski(1981)]{Dobaczewski1981123}
J.~Dobaczewski and J.~Skalski, \emph{Nucl. Phys. A} \textbf{369}, 123
  (1981).

\bibitem[Hinohara et~al.(2010)]{PhysRevC.82.064313}
N.~Hinohara, K.~Sato, T.~Nakatsukasa, M.~Matsuo, and K.~Matsuyanagi,
  \emph{Phys. Rev. C} \textbf{82}, 064313 (2010).

\bibitem[Matsuo et~al.(2000)]{PTP.103.959}
M.~Matsuo, T.~Nakatsukasa, and K.~Matsuyanagi, \emph{Prog. Theor. Phys.}
  \textbf{103}, 959 (2000).

\bibitem[Sato and Hinohara(2011)]{Sato201153}
K.~Sato and N.~Hinohara, \emph{Nucl. Phys. A} \textbf{849}, 53 (2011).

\bibitem[Stoitsov et~al.(2005)]{Stoitsov200543}
M.~Stoitsov, J.~Dobaczewski, W.~Nazarewicz, and P.~Ring, \emph{Comp. Phys.
  Comm.} \textbf{167}, 43  (2005).

\bibitem[Sakamoto and Kishimoto(1990)]{Sakamoto1990321}
H.~Sakamoto and T.~Kishimoto, \emph{Phys. Lett. B} \textbf{245}, 321
  (1990).

\bibitem[Deacon et~al.(2010)]{PhysRevC.82.034305}
A.~N. Deacon, {\it et al.},
\emph{Phys. Rev. C} \textbf{82}, 034305 (2010).

\bibitem[Takeuchi et~al.(2009)]{PhysRevC.79.054319}
S.~Takeuchi, {\it et al.}, \emph{Phys. Rev. C} \textbf{79}, 054319 (2009).

\bibitem[Yoneda et~al.(2001)]{Yoneda2001233}
K.~Yoneda, {\it et al.}, 
\emph{Phys. Lett. B} \textbf{499}, 233 (2001).

\bibitem[Gade et~al.(2007)]{gade:072502}
A.~Gade, {\it et al.}, \emph{Phys. Rev. Lett.} \textbf{99}, 072502 (2007).

\bibitem[Niedermaier et~al.(2005)]{PhysRevLett.94.172501}
O.~Niedermaier {\it et al.}, \emph{Phys. Rev. Lett.} \textbf{94}, 172501 (2005).

\bibitem[Motobayashi et~al.(1995)]{Motobayashi19959}
T.~Motobayashi, {\it et al.},
  \emph{Phys. Lett. B} \textbf{346}, 9 (1995).

\bibitem[Iwasaki et~al.(2001)]{Iwasaki2001227}
H.~Iwasaki, {\it et al.}, \emph{Phys. Lett. B} \textbf{522}, 227 (2001).

\bibitem[Hinohara et~al.(????)]{Mginprep}
N.~Hinohara, K.~Sato, K.~Yoshida, T.~Nakatsukasa, M.~Matsuo, and
  K.~Matsuyanagi, in preparation.

\end{thebibliography}


\end{document}